\documentclass[aps,prd,twocolumn,superscriptaddress,preprintnumbers,showpacs,altaffilletter]{revtex4}

\usepackage{graphicx}
\usepackage{dcolumn}
\usepackage{bm}
\usepackage{amssymb}

\def\aver#1{<\!\!#1\!\!>}

\begin{document}

\preprint{UASLP--IF--07--003}
\preprint{FERMILAB--Pub--07--312--E}

\title{
~\\~\\
~\\~\\
Polarization of \bm{$\Lambda^0$} and \bm{$\overline{\Lambda^0}$}
inclusively produced by 
\bm{$610\,\mbox{GeV}/c$} \bm{$\Sigma^-$} and 
\bm{$525\,\mbox{GeV}/c$} proton beams}


\affiliation{Ball State University, Muncie, IN 47306, U.S.A.}
\affiliation{Bogazici University, Bebek 80815 Istanbul, Turkey}
\affiliation{Carnegie-Mellon University, Pittsburgh, PA 15213, U.S.A.}
\affiliation{Centro Brasileiro de Pesquisas F\'{\i}sicas, Rio de Janeiro, Brazil}
\affiliation{Fermi National Accelerator Laboratory, Batavia, IL 60510, U.S.A.}
\affiliation{Institute for High Energy Physics, Protvino, Russia}
\affiliation{Institute of High Energy Physics, Beijing, P.R. China}
\affiliation{Institute of Theoretical and Experimental Physics, Moscow, Russia}
\affiliation{Max-Planck-Institut f\"ur Kernphysik, 69117 Heidelberg, Germany}
\affiliation{Moscow State University, Moscow, Russia}
\affiliation{Petersburg Nuclear Physics Institute, St.\ Petersburg, Russia}
\affiliation{Tel Aviv University, 69978 Ramat Aviv, Israel}
\affiliation{Universidad Aut\'onoma de San Luis Potos\'{\i}, San Luis Potos\'{\i}, Mexico}
\affiliation{Universidade Federal da Para\'{\i}ba, Para\'{\i}ba, Brazil}
\affiliation{University of Bristol, Bristol BS8~1TL, United Kingdom}
\affiliation{University of Iowa, Iowa City, IA 52242, U.S.A.}
\affiliation{University of Michigan-Flint, Flint, MI 48502, U.S.A.}
\affiliation{University of Rome ``La Sapienza'' and INFN, Rome, Italy}
\affiliation{University of S\~ao Paulo, S\~ao Paulo, Brazil}
\affiliation{University of Trieste and INFN, Trieste, Italy}
\author{J.L.~S\'anchez-L\'opez}
\affiliation{Universidad Aut\'onoma de San Luis Potos\'{\i}, San Luis Potos\'{\i}, Mexico}
\author{K.D.~Nelson}
\altaffiliation{Present address: University of Alabama at Birmingham, Birmingham, AL 35294}
\affiliation{University of Iowa, Iowa City, IA 52242, U.S.A.}
\author{J.~Engelfried}
\email{jurgen@ifisica.uaslp.mx}
\affiliation{Universidad Aut\'onoma de San Luis Potos\'{\i}, San Luis Potos\'{\i}, Mexico}
\author{U.~Akgun}
\affiliation{University of Iowa, Iowa City, IA 52242, U.S.A.}
\author{G.~Alkhazov}
\affiliation{Petersburg Nuclear Physics Institute, St.\ Petersburg, Russia}
\author{J.~Amaro-Reyes}
\affiliation{Universidad Aut\'onoma de San Luis Potos\'{\i}, San Luis Potos\'{\i}, Mexico}
\author{A.G.~Atamantchouk}
\altaffiliation{deceased}
\affiliation{Petersburg Nuclear Physics Institute, St.\ Petersburg, Russia}
\author{A.S.~Ayan}
\affiliation{University of Iowa, Iowa City, IA 52242, U.S.A.}
\author{M.Y.~Balatz}
\altaffiliation{deceased}
\affiliation{Institute of Theoretical and Experimental Physics, Moscow, Russia}
\author{A.~Blanco-Covarrubias}
\affiliation{Universidad Aut\'onoma de San Luis Potos\'{\i}, San Luis Potos\'{\i}, Mexico}
% ...
\collaboration{The SELEX Collaboration}
\noaffiliation

\date{June 23, 2007}

\begin{abstract}
We have measured the polarization of  $\Lambda^0$ and $\overline{\Lambda^0}$
inclusively produced by  $610\,\mbox{GeV}/c$ $\Sigma^-$ and
$525\,\mbox{GeV}/c$ proton beams in the
experiment SELEX during the 1996/7 fixed target run at Fermilab.
The polarization was measured as a function of 
the $\Lambda$ longitudinal momentum fraction $x_F$ 
and transverse momentum $p_t$.
For the $\Lambda^0$ produced by $\Sigma^-$ the polarization is
increasing with $x_F$, from slightly negative at $x_F\sim0$ to 
about $15\,\%$ at large $x_F$;
it shows a non-monotonic behavior as
a function of $p_t$.
For the proton beam, the $\Lambda^0$ polarization is negative and
decreasing as a function of $x_F$ and $p_t$.
The $\overline{\Lambda^0}$ polarization is compatible with $0$
for both beam particles over the full kinematic range.
The target dependence was examined but no statistically significant
difference was found.
\end{abstract}


\maketitle

\section{Introduction}
A large number of theoretical models have been constructed over the years
since it was first observed that inclusively produced hyperons are polarized.
These models have met with varying degrees of success, but it is clear
that more data and theoretical work are needed to clarify the picture.
A review of the current status is found 
in~\cite{Lach:1995jg,Felix:1997az,Felix:1999tf,Felix:2001kr}.
$\Lambda$ polarization is well studied with a proton beam, both in 
inclusive~\cite{Heller:1978ty} and
exclusive~\cite{Felix:1996kw,Felix:1997iq,Felix:2001zr} reactions.
For $\Sigma^-$ beam, only the WA89 experiment at CERN
reports polarization measurements
for $\Lambda^0$ and $\overline{\Lambda^0}$ as function of $p_t$ for one
average value of $x_F$~\cite{Adamovich:1994gy}, and for $\Lambda^0$ 
as function of both $p_t$ and $x_F$~\cite{Adamovich:2004cc}. In the latter
case the polarization is negative for small $x_F$, incrementing
to positive values with $x_F$, but with a non-monotonic 
behavior as a function of $p_t$ for different values of $x_F$.

In this study we exploit the capabilities of the SELEX apparatus to measure
the polarization of $\Lambda^0$ and $\overline{\Lambda^0}$
inclusively produced 
by $\Sigma^-$ and proton beams. 
The goal is to measure the polarizations as functions of $x_F$ and $p_t$
with higher beam momenta as the earlier WA89 measurement, using a different 
analysis method presenting different systematics, and extending the 
$\overline{\Lambda^0}$ polarization measurements to several values in $x_F$;
in addition, we add as a cross check the measurements with proton beam
within the same experiment.

The SELEX (E781) experiment at Fermilab is a fixed target experiment designed
primarily for high statistics studies of charmed baryons produced by a
charged hyperon beam incident on a segmented copper/carbon target.
However, the versatility of the apparatus allowed for the study of other
reactions~\cite{Pogodin:2003hx} in the same target.
In this work we study the polarization of inclusively produced $\Lambda^0$
and $\overline{\Lambda^0}$.
More specifically we determine the polarization and its $x_F$ and $p_t$
dependence. We follow the Basel Convention~\cite{Hughes:1960aaa}
for the sign of the polarization.

\section{Experiment}

The charged hyperon beam, which we use as a primary beam for $\Lambda$
production, was obtained in the Fermilab proton center beamline by steering
$800\,\mbox{GeV}/c$ protons from the
Tevatron onto a $1\,\mbox{mm} \times 2\,\mbox{mm}\times 40\,\mbox{cm}$
beryllium target located at the entrance of a $7.3\,\mbox{m}$, 
$3.5\,\mbox{T}$ hyperon magnet~\cite{Lach:2000vh}.
A curved channel in this magnet selected a beam of negative (positive) 
particles with a
mean momentum of $610\,\mbox{GeV}/c$ ($520\,\mbox{GeV}/c$) and a spread of
$\Delta p/p\approx 8\,\%$ HWHM.
The negative beam consisted of approximately equal parts 
$\Sigma^-$  and $\pi^-$  with a small admixture of 
$\Xi^-$ and $K^-$, while the positive beam contained about $92\,\%$ protons,
 in a total of $6\times10^5$ beam particles per second.
The targeting angle was set to zero degrees, thereby insuring an
unpolarized incident beam.

\begin{figure}
\includegraphics[angle=-90,width=0.45\textwidth,clip]{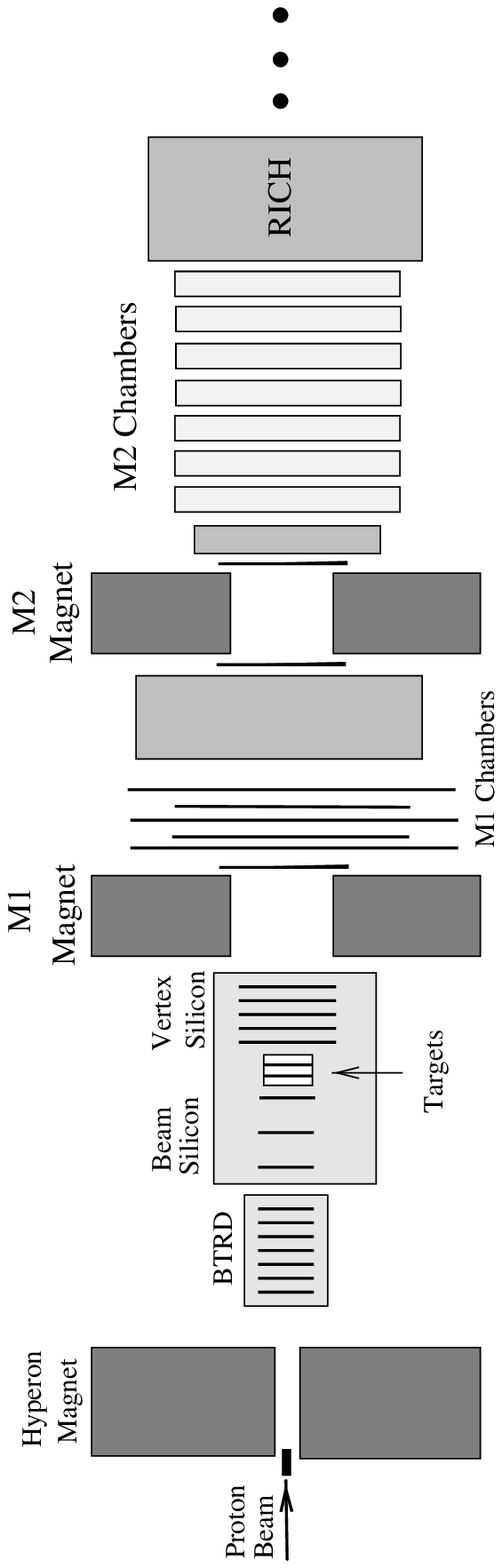}%
\caption{\label{figapp}
Experimental apparatus for this analysis (not to scale).}
\end{figure}

The portion of the experimental setup relevant to this analysis,
which is shown in Fig.~\ref{figapp}, consisted of a beam spectrometer,
a vertexing region and two downstream spectrometers (M1 and M2).
The beam spectrometer included the hyperon magnet, beam transition radiation
detectors (BTRD) and beam silicon strip detectors (BSSD). 
The vertex region
consisted of five segmented targets
(2~Cu, 3~C each separated by $1.5\,\mbox{cm}$ - altogether $5\,\%$ of an 
interaction length) and 20~planes of vertex silicon strip detectors (VSSD),
which resolved the interaction vertex and secondary vertices.
For VSSD tracks, the transverse position resolution was $4\,\mu\mbox{m}$
at $600\,\mbox{GeV}/c$. The M1 spectrometer was a wide angle spectrometer,
designed to analyze particles with momenta between 
$\sim2.5$ and $\sim15\,\mbox{GeV}/c$.
This spectrometer contained the M1~magnet ($\Delta p_t = 0.74\,\mbox{GeV}/c$),
large angle silicon strip detectors (LASD),
and proportional multi-wire chambers (PWC).
The second spectrometer (M2), downstream of M1, analyzed particles with 
momenta $\gtrsim15\,\mbox{GeV}/c$.
Its components were the M2 magnet ($\Delta p_t = 0.845\,\mbox{GeV}/c$),
LASD, PWC,
and a ring-imaging Cherenkov (RICH) detector~\cite{Engelfried:1998tv}
 used for particle identification, which provided 
$\pi/p$ separation up to $330\,\mbox{GeV}/c$. A hardware trigger
and software filter were used in SELEX to select the secondary interactions. 
The hardware trigger consisted of the beam, veto and hodoscope
scintillation counters.
The first level synchronized the trigger to beam particles.
The second level selected events where the BTRD determined the beam particle
was most likely a heavy particles ($\Sigma^-$ or proton), with an additional
refined definition in the analysis stage. 
The beam interacted in one of the five targets, and more than 3~charged tracks
were observed downstream of the target region of which at least 2
were positive tracks with momentum greater than $\sim15\,\mbox{GeV}/c$.
Triggered events accounted for $30\,\%$ of interactions.
A software filter looked for evidence of a secondary vertex and a
reconstructed primary vertex using high momentum ($\gtrsim15\,\mbox{GeV}/c$)
tracks. The software filter further reduced the data set by a factor of 8.
A sample of unfiltered data was always recorded for filter performance
and systematic error analysis.

\section{Data Analysis}
\begin{figure}
\includegraphics[width=0.45\textwidth]{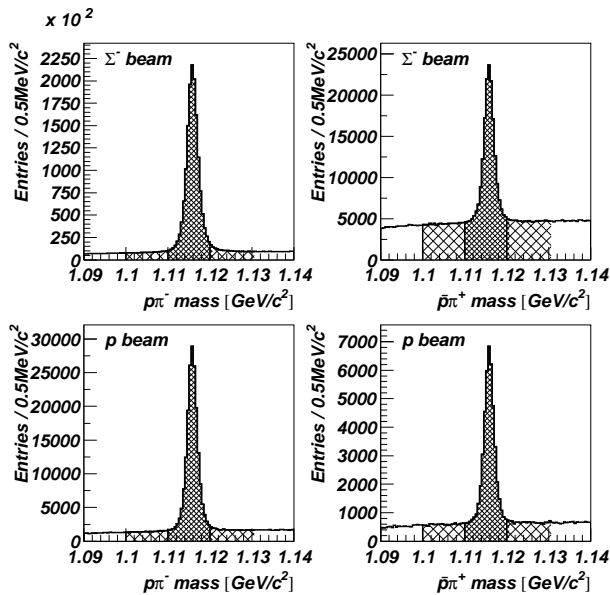}%
\caption{\label{signals}
Invariant mass distributions for $p\pi^-$ (left) and
$\overline{p}\pi^+$ (right) for $\Sigma^-$ (top) and proton (bottom) beams.
The signal and sideband regions are indicated.}
\end{figure}

The $\Lambda^0\to p\pi^-$ ($\overline{\Lambda^0}\to\overline{p}\pi^+$)
candidates were selected by
requiring oppositely
charge track pairs to form a vertex at least $5\,\sigma$ downstream of the
primary vertex, where $\sigma$ is the combined error of the $z$-coordinates
of the primary and secondary vertex, and upstream of the first VSSD plane.
The positive (negative) track was required to be identified by the RICH
as a proton candidate. In Fig.~\ref{signals} we show the invariant mass
distributions of these $p\pi^-$ and $\overline{p}\pi^+$ candidates, and
in table~\ref{tab:yields} we summarize the available statistics.
\begin{table}
\caption{\label{tab:yields} Number of $\Lambda^0$, 
$\overline{\Lambda^0}$ and $K_S^0$ for the different beams.}
\begin{ruledtabular}
\begin{tabular}{|c||r|r|r|r|}
Beam Particle & \multicolumn{1}{c|}{$\aver{p}$} &
\multicolumn{1}{c|}{\# $\Lambda^0$} & 
\multicolumn{1}{c|}{\# $\overline{\Lambda^0}$} & 
\multicolumn{1}{c|}{\# $K_S^0$} \\ \hline
$\Sigma^-$    & $611\,\mbox{GeV}/c$~ & 1,360,000 & 112,000&  4,698,000\\ \hline
$p$           & $525\,\mbox{GeV}/c$~ &   162,000 &  35,700 & 752,000\\
\end{tabular}
\end{ruledtabular}
\end{table}
The $K_S^0\to\pi^+\pi^-$ decays are also included to be able to cross check
the analysis procedure.

The polarization analysis consisted
of extracting the polarization $\bf{P}$ from a fit to the 
proton decay distribution in the $\Lambda$ rest frame:
\begin{equation}
\frac{dN_{\rm meas}}{d\cos{\theta}} \propto
A(\cos{\theta},x_F,p_t)[1+\alpha_{\Lambda}\bf{P}\cos{\theta}]
\label{eq:pol}
\end{equation}
where the $y$-axis is normal to the production plane (defined by the
vector of the incoming beam particle and the outgoing $\Lambda$),
the $z$-axis  is the 
direction of the $\Lambda$ line-of-flight, and
the $x$-axis completes the orthogonal triad.
$N_{\rm meas}$ is the number of events, $\theta$ is the angle between
the proton line of flight and the coordinate axes,
$\alpha_{\Lambda}=0.642\pm0.013$ is
the asymmetry parameter~\cite{Yao:2006px},
and $A$ is the acceptance function.
By parity conservation in the production process,
the only polarization allowed is along the y-axis.
Clearly, no polarization should be measured along the
$x$ or $z$ direction.

Apparatus (or false) asymmetries are present along all axes.
The false asymmetries can be removed by using a bias canceling technique,
or by modeling the acceptance function $A$ with Monte Carlo simulations. Bias
canceling requires a subdivision of the available statistics, while the
Monte Carlo simulation requires a good understanding of the apparatus and
extensive computing time.  Due to the overall low statistics compared to
other measurements we present here results
obtained~\cite{Sanchez-Lopez:2006kn} with the latter; however,
with the bias canceling technique we obtained compatible
results~\cite{Nelson:1999kn}.
The analysis methods were validated by a non-observation of polarization
in the forbidden projections, by analyzing the $K_S^0$ with the same programs,
and by Monte Carlo simulation via re-obtaining a known polarization.
\begin{table*}
\caption{\label{tab:lamsig} 
Polarization (in \%) of $\Lambda^0$ inclusively 
produced by $\Sigma^-$ as a function
of $x_F$ and $p_t$. The same information is shown in 
figures~\ref{fig:lamsigxf} and \ref{fig:lamsigpt}.}
\begin{ruledtabular}
\begin{tabular}{|c|cccccc|}
 & \multicolumn{6}{c|}{$\aver{p_T}~~[~\mbox{GeV}/c~]$}  \\ 
$x_F$  & 0.1 & 0.3 & 0.5   & 0.7 & 0.9 & 1.1  \\ 
\hline
0.15 &  $1.3\pm  1.7$& $-0.5\pm  1.2$& $-3.5\pm  1.1$&
        $1.9\pm  1.2$&  $0.4\pm  1.5$& $-1.6\pm  1.9$\\
0.25 & $0.4\pm  2.0$& $-0.4\pm  1.4$&  $1.0\pm  1.4$&
     $0.2\pm  1.6$& $-1.0\pm  1.9$& $-0.8\pm  2.6$\\
0.35 &$3.5\pm  2.6$& $-0.6\pm  1.8$&  $1.7\pm  1.8$&
       $7.5\pm  1.9$&  $4.7\pm  2.4$&  $0.7\pm  3.3$\\
0.45 &$2.6\pm  3.4$&  $8.1\pm  2.3$&  $9.9\pm  2.3$&
       $6.4\pm  2.5$&  $6.6\pm  3.0$& $10.8\pm  4.2$\\
0.55 &$5.4\pm  5.2$&  $5.4\pm  3.3$& $10.5\pm  3.2$&
      $12.6\pm  3.4$& $10.0\pm  4.2$&  $6.3\pm  5.7$\\
0.65 &$7.0\pm  7.7$&  $5.4\pm  4.7$& $22.6\pm  4.7$&
      $16.4\pm  5.4$& $11.2\pm  6.4$& $21.4\pm  8.8$\\
0.75 &$-8.5\pm 10.3$& $11.8\pm  7.3$& $13.6\pm  7.5$&
      $13.0\pm  9.3$& $-9\pm 13$&  $5\pm 17$\\
0.85 &$25\pm 19$& $16\pm 14$& $11\pm 16$ &
      $41\pm 22$& $10\pm 25$& $56\pm 53$\\
\end{tabular}
\end{ruledtabular}
\end{table*}

To obtain the polarization of $\Lambda^0$ produced by a $\Sigma^-$ beam,
we divided the data into 100~bins, 10 each in $0.0\le x_F\le1.0$, and in
$0.0\,\mbox{GeV}/c\le p_t \le 2.0\,\mbox{GeV}/c$. The dependence
of the acceptance function~$A$ on $x_F$ and $p_t$ can be neglected
within one bin, and only the dependence on $\cos\theta$ has to be taken into
account.  For each $[x_F,p_t]$ bin a two-dimensional histogram of the cosine
of the angle versus the invariant mass of the $p\pi^-$ is filled and the
$dN_{\rm meas}/d\cos\theta$ distribution is obtained via sideband
subtraction (see also Fig.~\ref{signals}). 
To correct for the acceptance, we obtain the same distribution
from Monte Carlo simulation (we verified that the acceptance is independent
of an initial polarization of the Monte Carlo sample), and correct for 
the acceptance in each $[x_F,p_t]$ bin separately. 
A straight line fit is performed to the final distribution,
and the polarization is extracted according to equation~\ref{eq:pol}.

For the lower statistics sample of $\overline{\Lambda^0}$
and the proton beam data, it was not possible to subdivide the available
data as before. Only measurements as a function of $x_F$ and $p_t$, 
averaging over the other variable, could be performed.  As described
above, a two-dimensional histogram was filled, but this time with a weight 
factor obtained from an acceptance model which takes into account the
$x_F$, $p_t$ as well as the primary interaction target dependence of
the acceptance~\cite{Engelfried:2006aab}.

\section{Systematic Checks}

We performed several systematic checks to verify our analysis method,
including the algorithm as well as the acceptance model.
We analysed simulated
samples of known polarization, and always re-obtained the correct result.
We  observed asymmetries in forbidden
projections compatible with $0$.
We also measured the asymmetry of $K_S^0\to\pi^+\pi^-$ decays
which again is compatible with $0$ over the full kinematic range as shown
in the last columns of tables~\ref{tab:otherspt} and~\ref{tab:othersxf}.

Other systematics checks included harder cuts on the separation between primary
vertex and $\Lambda$ decay vertex, requiring that also the pion track reaches
the M2 spectrometer, and that the pion is identified with the RICH
detector. We always obtained, within the statistical errors, the same
results for all polarizations. For these reasons we only quote statistical
errors on all our measurements.

\section{Results}
\begin{figure}
\includegraphics[width=0.45\textwidth]{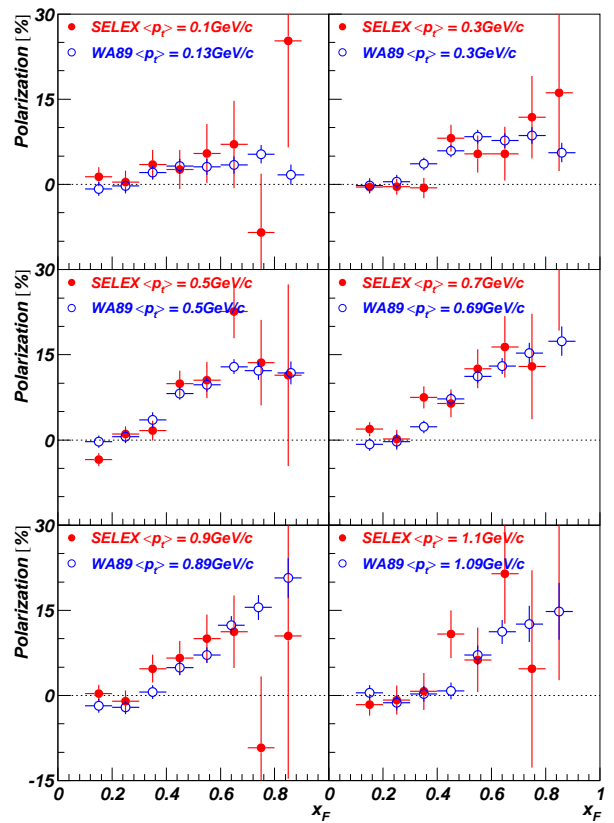}%
\caption{\label{fig:lamsigxf}
Polarization of $\Lambda^0$ inclusively produced by $\Sigma^-$ as a function
of $x_F$ for different $p_t$ values. Also shown are data from
ref.~\cite{Adamovich:2004cc}.  The SELEX data points are also given in
table~\ref{tab:lamsig}.}
\end{figure}

\begin{figure}
\includegraphics[width=0.45\textwidth]{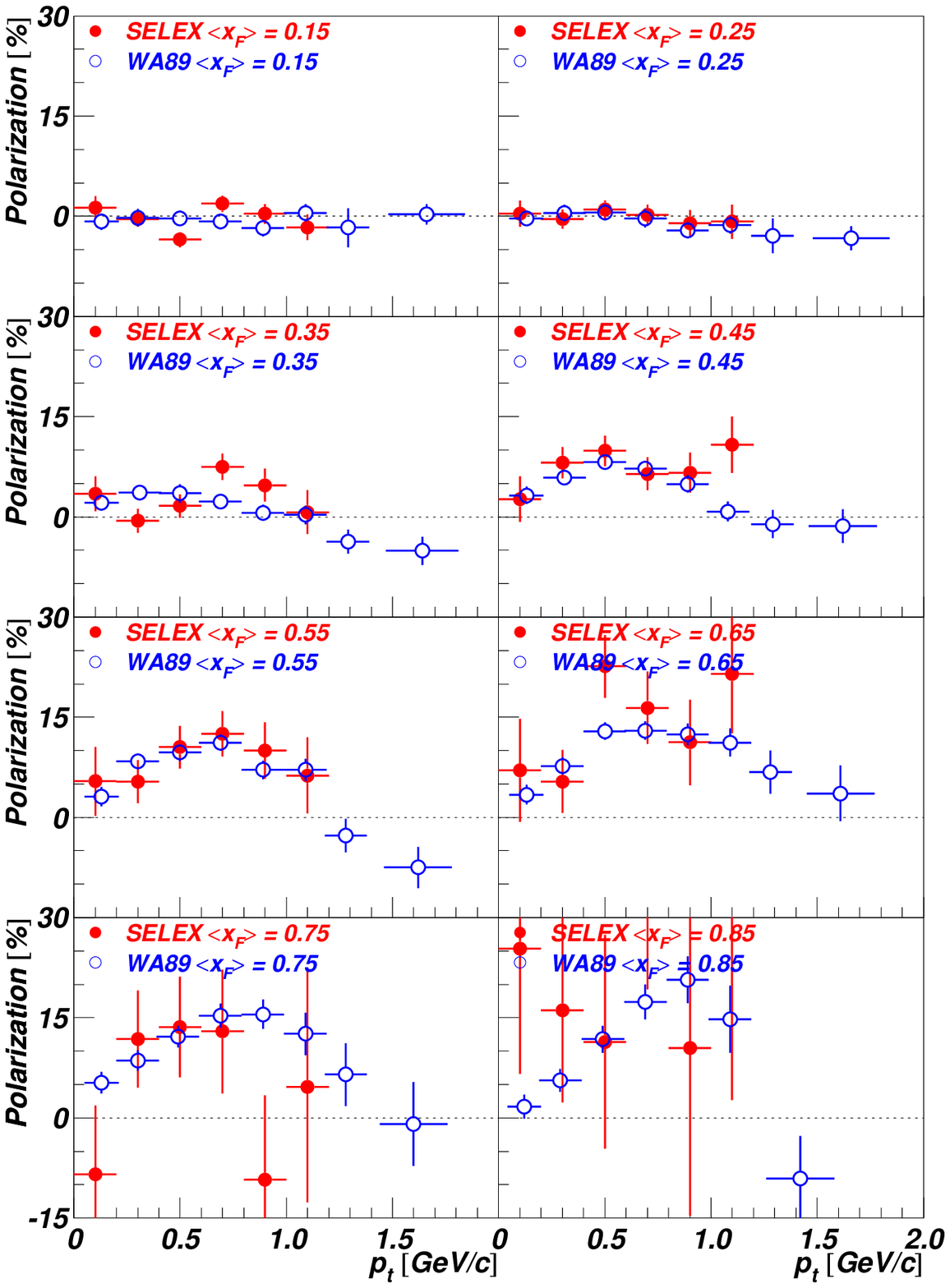}%
\caption{\label{fig:lamsigpt}
Polarization of $\Lambda^0$ inclusively produced by $\Sigma^-$ as a function
of $p_t$ for different $x_F$ values. Also shown are data from
ref.~\cite{Adamovich:2004cc}.  The SELEX data points are also given in
table~\ref{tab:lamsig}.}
\end{figure}

In Figs.~\ref{fig:lamsigxf} and \ref{fig:lamsigpt}, as well as
in table~\ref{tab:lamsig} we present our
results for the $\Lambda^0$ polarization with a $\Sigma^-$ beam
as function of $x_F$ and $p_t$.

\begin{figure}
\includegraphics[width=0.45\textwidth]{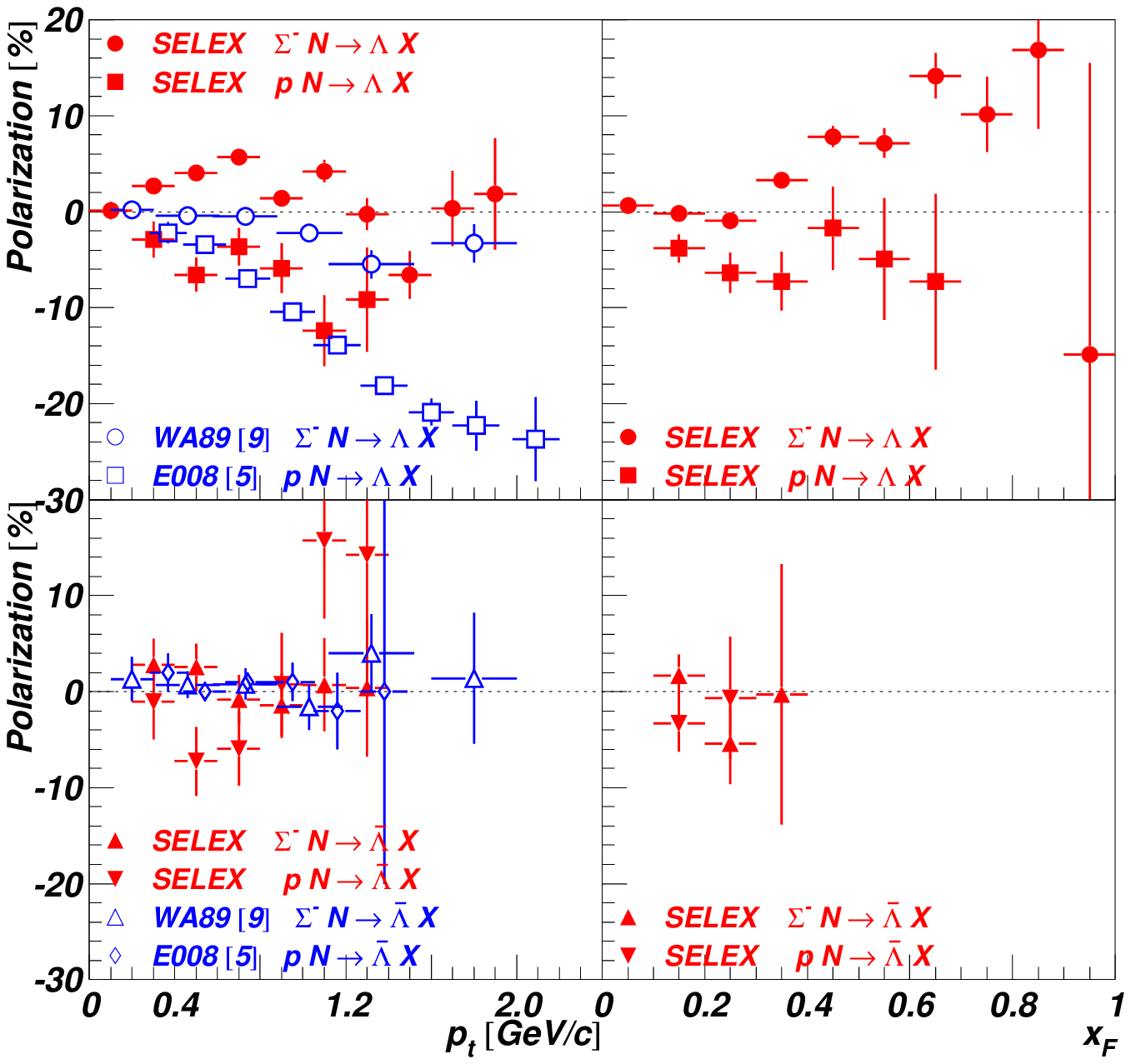}%
\caption{\label{fig:others}Polarization of $\Lambda^0$ (top) and
$\overline{\Lambda^0}$ (bottom)
produced by $\Sigma^-$ and protons,
as function of $p_t$ (left) and $x_F$ (right).
Also shown are data from
Refs.~\cite{Heller:1978ty,Adamovich:1994gy}.
The SELEX data points are also given in tables~\ref{tab:otherspt} and
\ref{tab:othersxf}.}
\end{figure}

Figure~\ref{fig:others} shows the polarization of $\Lambda^0$ and
$\overline{\Lambda^0}$ 
produced by $\Sigma^-$ and protons.
Due to the lower statistics available, only the distributions for an
average value of $x_F$ and $p_t$ are shown.  The same information is
presented in tables~\ref{tab:otherspt} and \ref{tab:othersxf}.
\begin{table*}
\caption{\label{tab:otherspt}Polarization (in \%) of $\Lambda^0$ and
$\overline{\Lambda^0}$ 
produced by $\Sigma^-$ and protons, as
function of $p_t$, averaged over $x_F$.
 The same information is presented graphically in
Fig.~\ref{fig:others} (left).
Also shown are the asymmetry values measured for the $K_S^0$.}
\begin{ruledtabular}
\begin{tabular}{|c| c c c c c|}
&
$pN\to\Lambda^0X$ &
$pN\to\overline{\Lambda^0}X$ & 
$\Sigma^-N\to\overline{\Lambda^0}X$ &
$\Sigma^-N\to\Lambda^0X$ &
$\Sigma^-N\to K_S^0X$\\ \cline{1-1}
$p_t$ $[\mbox{GeV}/c]$& $\aver{x_F}=0.24$ &
$\aver{x_F}=0.11$ &
$\aver{x_F}=0.11$ &
$\aver{x_F}=0.29$ &
Asymmetry\\
\hline
  $0.1$& -- & -- & -- &  
  $ 0.1\pm 0.9$ &
  $-0.1\pm 0.4$ \\
  $0.3$&  $-2.9\pm1.9$ &
  $-1.0 \pm   3.9$ &
  $ 2.8 \pm   2.8$ &
  $ 2.7 \pm   0.6$ & 
  $ 0.6 \pm   0.3$ \\
  $0.5$&  $-6.6\pm1.8$ &
  $-7.3 \pm   3.6$ &
  $ 2.5 \pm   2.4$ &
  $ 4.0 \pm   0.6$ &
  $-0.1 \pm   0.3$ \\
  $0.7$&  $-3.7\pm2.0$ &
  $-6.0 \pm   3.9$ &
  $-0.9 \pm   2.6$ &
  $ 5.7 \pm   0.7$ &
  $ 0.3 \pm   0.4$ \\
  $0.9$&  $-5.9\pm2.6$ &
  $ 0.8 \pm   5.4$ &
  $-1.5 \pm   3.3$ &
  $ 1.4 \pm   0.9$ &
  $ 0.6 \pm   0.5$ \\
  $1.1$& $-12.4\pm3.7$ &
  $15.8 \pm   8.2$ &
  $ 0.7 \pm   4.9$ &
  $ 4.2 \pm   1.2$ &
  $-1.0 \pm   0.7$ \\
  $1.3$& $ -9.2\pm5.4$ &
  $14.3 \pm  13.3$ &
  $ 0.4 \pm   7.2$ &
  $-0.3 \pm   1.7$ &
  $ 0.3 \pm   1.0$  \\
  $1.5$& -- & -- & -- &  
  $-6.6 \pm   2.5$ &
  $ 0.6 \pm   1.5$ \\
  $1.7$& -- & -- & -- &  
  $ 0.3 \pm   3.9$ &
  $-2.1 \pm   2.3$\\
  $1.9$& -- & -- & -- &  
  $ 1.9 \pm   5.8$ &
  $-2.8 \pm   3.5$
\end{tabular}
\end{ruledtabular}
\end{table*}
\begin{table*}
\caption{\label{tab:othersxf}
Polarization (in \%) of $\Lambda^0$ and
$\overline{\Lambda^0}$ 
produced by $\Sigma^-$ and protons, as
function of $x_F$, averaged over $p_t$.
 The same information is presented graphically in
Fig.~\ref{fig:others} (right).
Also shown are the asymmetry values measured for the $K_S^0$.}
\begin{ruledtabular}
\begin{tabular}{|c| c c c c c|}
$\aver{p_t}$&
$pN\to\Lambda^0X$ &
$pN\to\overline{\Lambda^0}X$ &
$\Sigma^-N\to\overline{\Lambda^0}X$ &
$\Sigma^-N\to\Lambda^0X$ &
$\Sigma^-N\to K_S^0X$\\ \cline{1-1}
$x_F$ & $0.59\,\mbox{GeV}/c$ &
$0.60\,\mbox{GeV}/c$ &
$0.63\,\mbox{GeV}/c$ &
$0.57\,\mbox{GeV}/c$ &
Asymmetry\\
\hline
  $0.05$& -- & -- & -- &  
  $0.6 \pm  0.6$ &
  $0.4 \pm  0.2$ \\
 $0.15$&  $-3.8\pm1.5$ &
 $ -3.3 \pm   3.0$ &
 $  1.7 \pm   2.2$ &
 $ -0.2 \pm  0.5$ &
 $  0.0 \pm  0.3$ \\
 $0.25$&  $-6.4\pm2.1$ &
 $ -0.7  \pm  6.4$ &
 $ -5.4 \pm   4.3$ &
 $  -1.0 \pm  0.7$ &
 $  -0.3 \pm  0.5$ \\
 $0.35$&  $-7.2\pm3.1$ &
 $ 42 \pm  15$ &
 $ 0 \pm  14$ &
 $   3.3 \pm  0.8$ &
 $  -0.4 \pm  1.0$ \\
 $0.45$&  $-1.7\pm4.4$ &
 $-23 \pm  36$ &
 $ -29 \pm  50$ &
 $    7.8 \pm  1.1$ &
 $    1.4 \pm  2.3$ \\
 $0.55$&  $-5.0\pm6.3$ &
     -- &
 $  25 \pm  61$ &
 $   7.2 \pm  1.6$ &
 $  -4.4 \pm  5.2$  \\
 $0.65$&  $-7.3\pm9.2$&  
     -- &
     -- &
 $  14.2 \pm  2.4$ &
 $   7 \pm 14$ \\
  $0.75$& -- & -- & -- &  
  $  10.1 \pm  3.9$ &
  $  48 \pm 46$ \\
  $0.85$& -- & -- & -- &  
  $  16.9 \pm  8.2$ &
     -- \\
  $0.95$& -- & -- & -- &  
  $  -15 \pm 30$ &
       --
\end{tabular}
\end{ruledtabular}
\end{table*}

\section{Discussion and Conclusions}
For the inclusive production of $\Lambda^0$ by a $\Sigma^-$ beam, our results
confirm the WA89~\cite{Adamovich:1994gy,Adamovich:2004cc} measurements
(obtained via the bias canceling method) of 
a generally positive polarization, increasing with $x_F$, as shown
in Fig.~\ref{fig:lamsigxf} and \ref{fig:lamsigpt}. 
At small $p_t$ and $x_f$ the polarization is almost $0$, and in general
the dependence on $p_t$ is non-monotonic for different bins of $x_F$,
increasing and decreasing after reaching some maximum value.
Comparing to the earlier, lower-statistics 
WA89~\cite{Adamovich:1994gy} result 
with the dependence on $p_t$ for
$\aver{x_F}=0.30$
(Fig.~\ref{fig:others}, upper left and table~\ref{tab:otherspt}), 
we observe small polarization values confirming the previous
measurement. The authors of~\cite{Adamovich:2004cc} attribute the non-monotonic
behavior of the polarization as function of $p_t$ and $x_F$ to different
production mechanism in different kinematic regimes.
We also examined the target dependence, but no difference was found.

For the polarization for $\Lambda^0$ produced by a proton beam as a function
of $p_t$, our results
coincide with the higher statistics, but lower beam momentum,
 results from E008~\cite{Heller:1978ty}.
In Fig.~\ref{fig:others} (upper right) and
table~\ref{tab:othersxf} we also present
polarization results as a function of $x_F$ and compare them to our $\Sigma^-$
data.

The $\overline{\Lambda^0}$ polarization is compatible with $0$, for 
$\Sigma^-$ and proton beams for all $p_t$ values measured, 
as
obtained in~\cite{Heller:1978ty,Adamovich:1994gy}.  In addition we present
also the polarization as function of
$x_F$ (Fig.~\ref{fig:others} (lower right) and
table~\ref{tab:othersxf}).

In conclusion, we measured the polarization of inclusively produced
$\Lambda^0$
with a $\Sigma^-$ beam, and confirming the WA89 results 
with a different analysis method
and higher beam momentum.
We do not observe any polarization of the  $\overline{\Lambda^0}$ at
$p_t\le1.3\,\mbox{GeV}/c$ and $x_F\le0.55$.
The $\Lambda^0$ polarization with a proton beam is compatible with earlier
results.

\begin{acknowledgments}
The authors are indebted to the staff of Fermi National Accelerator Laboratory
and for invaluable technical support from the staffs of collaborating
institutions.
This project was supported in part by Bundesministerium f\"ur Bildung, 
Wissenschaft, Forschung und Technologie, Consejo Nacional de 
Ciencia y Tecnolog\'{\i}a {\nobreak (CONACyT)},
the Secretar\'{\i}a de Educaci\'on P\'ublica (Mexico)
(grant number 2003-24-001-026),
Fondo de Apoyo a la Investigaci\'on (UASLP),
Conselho Nacional de Desenvolvimento Cient\'{\i}fico e Tecnol\'ogico,
Funda\c{c}\~ao de Amparo \`a Pesquisa do Estado de S\~ao Paulo (FAPESP),
the Israel Science Foundation founded by the Israel Academy of Sciences and 
Humanities, Istituto Nazionale di Fisica Nucleare (INFN),
the International Science Foundation (ISF),
the National Science Foundation (Phy \#9602178),
NATO (grant CR6.941058-1360/94),
the Russian Academy of Science,
the Russian Ministry of Science and Technology,
the Russian Foundation for Basic Research (RFBR grant 05-02-17869),
the Turkish Scientific and Technological Research Board (T\"{U}B\.ITAK),
and the U.S.\ Department of Energy (DOE grant DE-FG02-91ER40664 and
DOE contract number DE-AC02-76CHO3000).\\
~~~~~~~~~~~~~~~~~~~~~~~~~~~~~~~~~~~~~~~

~~~~~~~~~~~~~~~~

~~~~~~~~~~~~~~~

~~~~~~~~~~~~~~

\end{acknowledgments}

\end{document}